\documentclass[a4paper,12pt]{article}
\usepackage{amssymb,amsmath,amsfonts}
\usepackage{fullpage}
\usepackage[latin1]{inputenc}
\usepackage[dvips]{graphicx}

\begin{document}
\title{The Beta Generalized Exponential Distribution}
\date{}
\author{Wagner Barreto-Souza$^{\rm a}$, Alessandro H. S. Santos$^{\rm b}$ and Gauss M. 
Cordeiro$^{\rm b}$}

\author{Wagner Barreto-Souza$^{\rm a}$, Alessandro H. S. Santos$^{\rm b}$ and Gauss M. 
Cordeiro$^{\rm b}$\\\\
$^{\rm a}${\em Departamento de Estat\'{\i}stica},\\
Universidade Federal de Pernambuco,\\
Cidade Universitária, 50740-540 -- Recife, PE, Brazil\\
(wagnerbs85@hotmail.com); \\\\
$^{\rm b}${ \em Departamento de Estat\'{\i}stica e Inform\'atica},\\
Universidade Federal Rural de Pernambuco,\\
Rua Dom Manoel de Medeiros s/n, 50171-900 -- Recife, PE, Brazil \\
(alessandrohss@yahoo.com.br, gausscordeiro@uol.com.br)}

\maketitle

\begin{abstract}
We introduce the beta generalized exponential distribution that includes 
the beta exponential and generalized exponential distributions as
special cases. We provide a comprehensive mathematical treatment of
this distribution. We derive the moment generating function and 
the $r$th moment thus generalizing some results in the literature. 
Expressions for the density, moment generating function and $r$th 
moment of the order statistics also are obtained. We discuss estimation of 
the parameters by maximum likelihood and provide the information matrix. 
We observe in one application to real data set that this model is quite flexible
and can be used quite effectively in analyzing positive data in
place of the beta exponential and generalized exponential distributions.\\

{\it\bf keywords}: Beta exponential distribution, Information matrix, Generalized
exponential dis\-tri\-bu\-tion, Maximum likelihood estimation.
\end{abstract}

\section{Introduction}

Gupta and Kundu \cite{Gupta1999} defined the cumulative distribution function (cdf) of the generalized
exponential (GE) distribution by
\begin{equation}\label{cd_ee}
G_{\lambda,\alpha}(x)=(1-e^{-\lambda x})^\alpha, \quad x>0.
\end{equation}
The two parameters of the GE distribution represent the shape ($\alpha>0$) and the scale parameter
($\lambda>0$) like the gamma and Weibull distributions. The distribution (\ref{cd_ee}) is also
named the exponentiated exponential distribution. Clearly, the exponential distribution is a particular
case of the GE distribution when $\alpha=1$. Gupta and Kundu also investigated some of their properties.
The exponentiated Weibull (EW) distribution introduced by Mudholkar et al. \cite{Mudholkar1993} extends the
GE distribution and was also studied by Mudholkar et al. \cite{Mudholkar1995}, Mudholkar and Hutson \cite{Mudholkar1996} 
and Nassar and Eissa \cite{Nassar2003}. Nadarajah and Kotz \cite{Nadarajah2006} introduced four more exponentiated
type distributions: the exponentiated gamma, exponentiated Weibull, ex\-po\-nen\-tia\-ted Gumbel and
exponentiated Fréchet distributions by generalizing the gamma, Weibull, Gumbel and Fréchet 
distributions in the same way that the GE distribution extends the exponential distribution. They also provide
some mathematical properties for each exponentiated distribution.\\

The GE density function varies significantly depending on the shape parameter $\alpha$. Also, the
hazard function is a non-decreasing function if $\alpha > 1$, and it is a non-increasing function if $\alpha < 1$.
For $\alpha=1$, it is constant. The GE distribution has lots of properties which are quite similar in nature
to those of the gamma distribution but it has explicit expressions for the distribution and survival functions
like a Weibull distribution. The gamma, Weibull and GE distributions extend the exponential
distribution but in different ways. Therefore, it can be used as an alternative to the Weibull and
gamma distributions and in some situations it might work better in terms of fitting than the other two
distributions although it can not be guaranteed. Moreover, it is well known that the
gamma distribution has certain advantages compared to the Weibull in terms of the
faster convergence of the maximum likelihood estimates (MLEs). It is expected 
that the GE distribution also should enjoy those properties.

Consider starting from the cdf $G(x)$ of a random variable, Eugene et al. \cite{Eugene2002}
defined a class of generalized distributions from it given by
\begin{equation}\label{beta_f}
F(x)=\frac{1}{B(a,b)}\int_0^{G(x)}\omega^{a-1}(1-\omega)^{b-1}d\omega,
\end{equation}
where $a>0$ and $b>0$ are two additional parameters whose role is to introduce
skewness and to vary tail weight and $B(a,b)=\int_0^1 \omega^{a-1}(1-\omega)^{b-1}d\omega$
is the beta function. The cdf $G(x)$ could be quite arbitrary and $F$ is named
the beta $G$ distribution. Application of $X=G^{-1}(V)$ to $V\sim B(a,b)$ yields
$X$ with cdf (\ref{beta_f}).

Eugene et al. \cite{Eugene2002} defined the beta normal (BN) distribution
by taking $G(x)$ in (\ref{beta_f}) to be the cdf of the normal distribution and derived 
some first moments. Ge\-ne\-ral expressions for the moments of the BN distribution were 
derived by Gupta and Nadarajah \cite{Gupta2004}. Nadarajah and Kotz \cite{Nadarajah2004} 
considered the beta Gumbel (BG) distribution by taking $G(x)$ as the cdf of the Gumbel
distribution and provided closed-form expressions for the moments,
the asymptotic distribution of the extreme order statistics and
discussed the maximum likelihood estimation procedure. Also,
Nadarajah and Kotz \cite{Nadarajah2005} worked with the beta
exponential (BE) distribution and obtained the moment generating
function, the first four cumulants, the asymptotic distribution of
the extreme order statistics and discussed the maximum likelihood
estimation. Some of the Nadarajah and Kotz's \cite{Nadarajah2005}
results were ge\-ne\-ra\-li\-zed by Cordeiro et al. \cite{Cordeiro2008} who
considered the beta Weibull distribution. They derived the moment
generating function, the moments and the information matrix.

We can write (\ref{beta_f}) by
\begin{equation}\label{cdf_*A}
F(x)= I_{G(x)}(a,b),
\end{equation}
where $I_{y}(a,b) = B(a,b)^{-1}\int_{0}^{y}w^{a-1}(1-w)^{b-1}dw$ denotes the incomplete beta
function ratio, i.e., the cdf of the beta distribution with parameters $a$ and $b$.
For general $a$ and $b$, we can express (\ref{cdf_*A}) in terms of the well-known hypergeometric
function defined by
\begin{equation*}\label{ghf}
_2F_{1}(\alpha,\beta,\gamma;x)= \sum_{i = 0}^{\infty}\frac{(\alpha)_{i}(\beta)_{i}}{(\gamma)_{i}i!}\,x^{i},
\end{equation*}
where $(\alpha)_{i}=\alpha (\alpha+1) \ldots (\alpha+i-1)$ denotes the ascending factorial. We obtain
\begin{equation*}\label{cdf_*Atrans}
F(x)= \frac{G(x)}{a\,B(a,b)}\,_2F_{1}(a,1-b,a+1;G(x)).
\end{equation*}
The properties of $F(x)$ for any beta $G$ distribution defined from a parent
$G(x)$ in (\ref{beta_f}) could, in principle, follow from the properties of the
hypergeometric function which are well established in the literature; see, for example,
Section 9.1 of Gradshteyn and Ryzhik \cite{Gradshteyn}.

The probability density function (pdf) corresponding to (\ref{beta_f})
can be put in the form
\begin{equation}\label{beta1_f}
f(x)=\frac{1}{B(a,b)}G(x)^{a-1}\{1-G(x)\}^{b-1} g(x),
\end{equation}
where we noted that $f(x)$ will be most tractable when the cdf $G(x)$ and the
pdf $g(x)=dG(x)/dx$ have simple analytic expressions. Except for some special choices for $G(x)$
in (\ref{beta_f}) as is the case when $G(x)$ is given by (\ref{cd_ee}), it would appear that 
the pdf $f(x)$ will be difficult to deal with.

\begin{figure}
\centering
\includegraphics[scale=0.45]{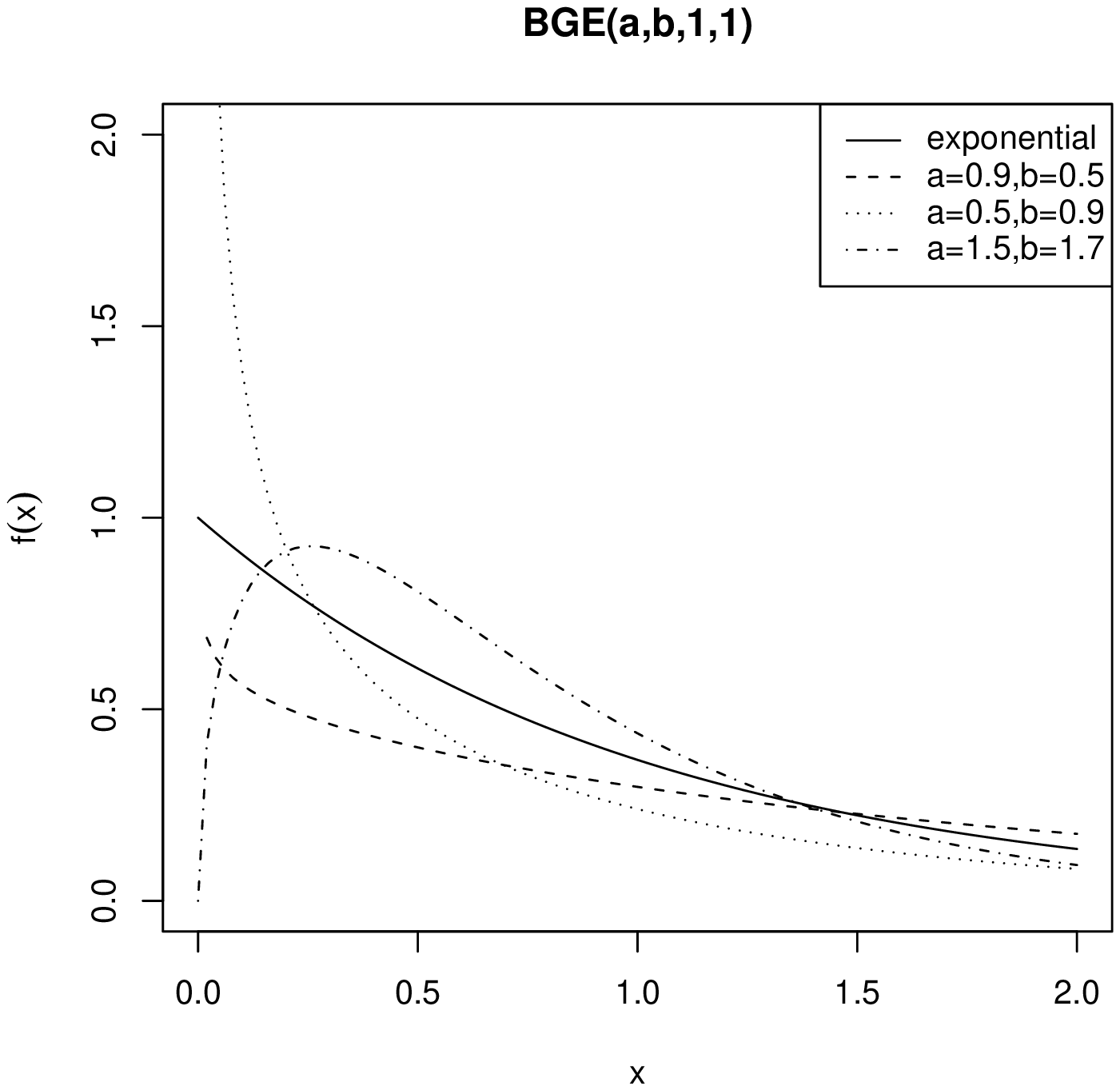}\includegraphics[scale=0.45]{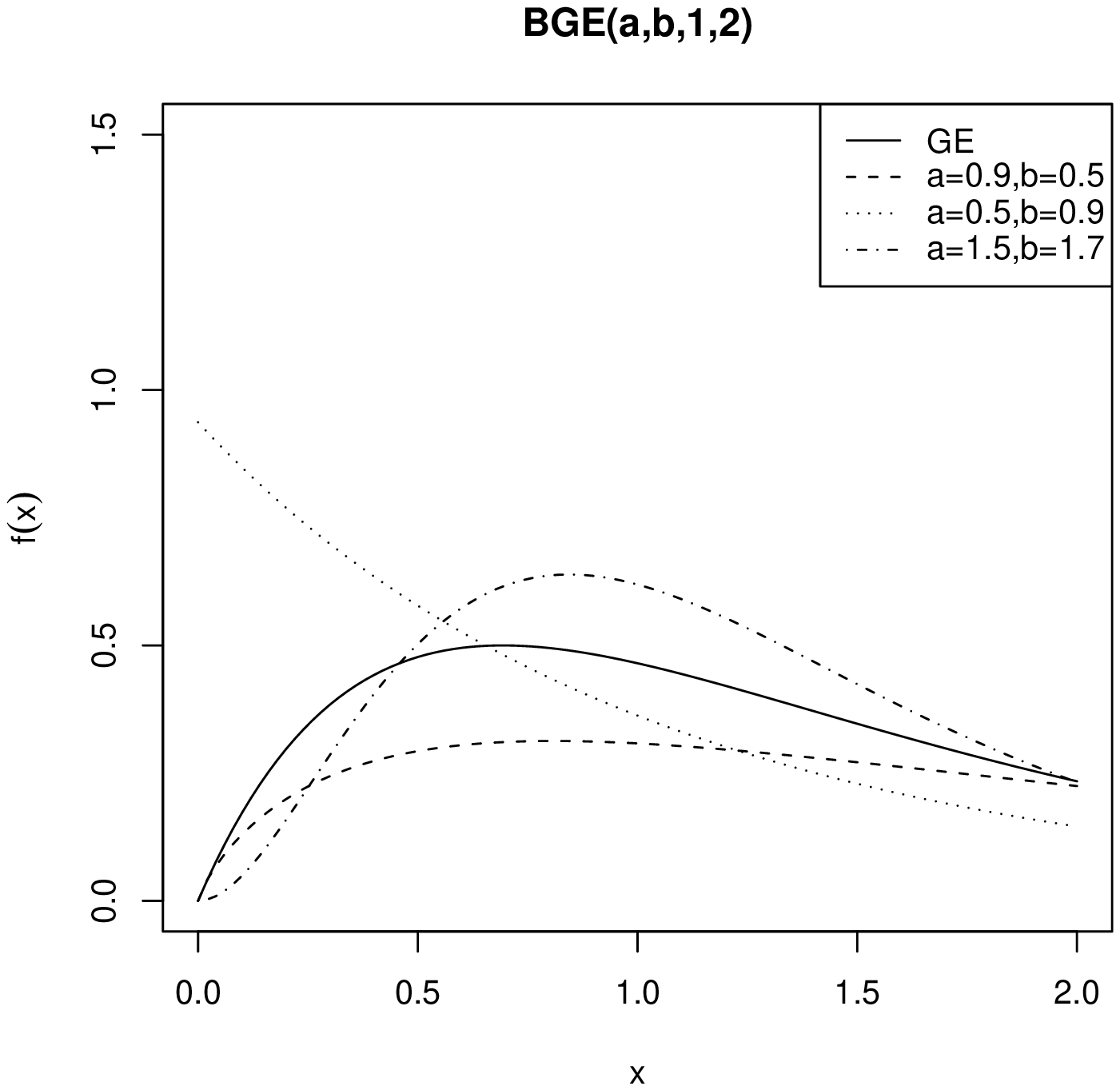}
\includegraphics[scale=0.45]{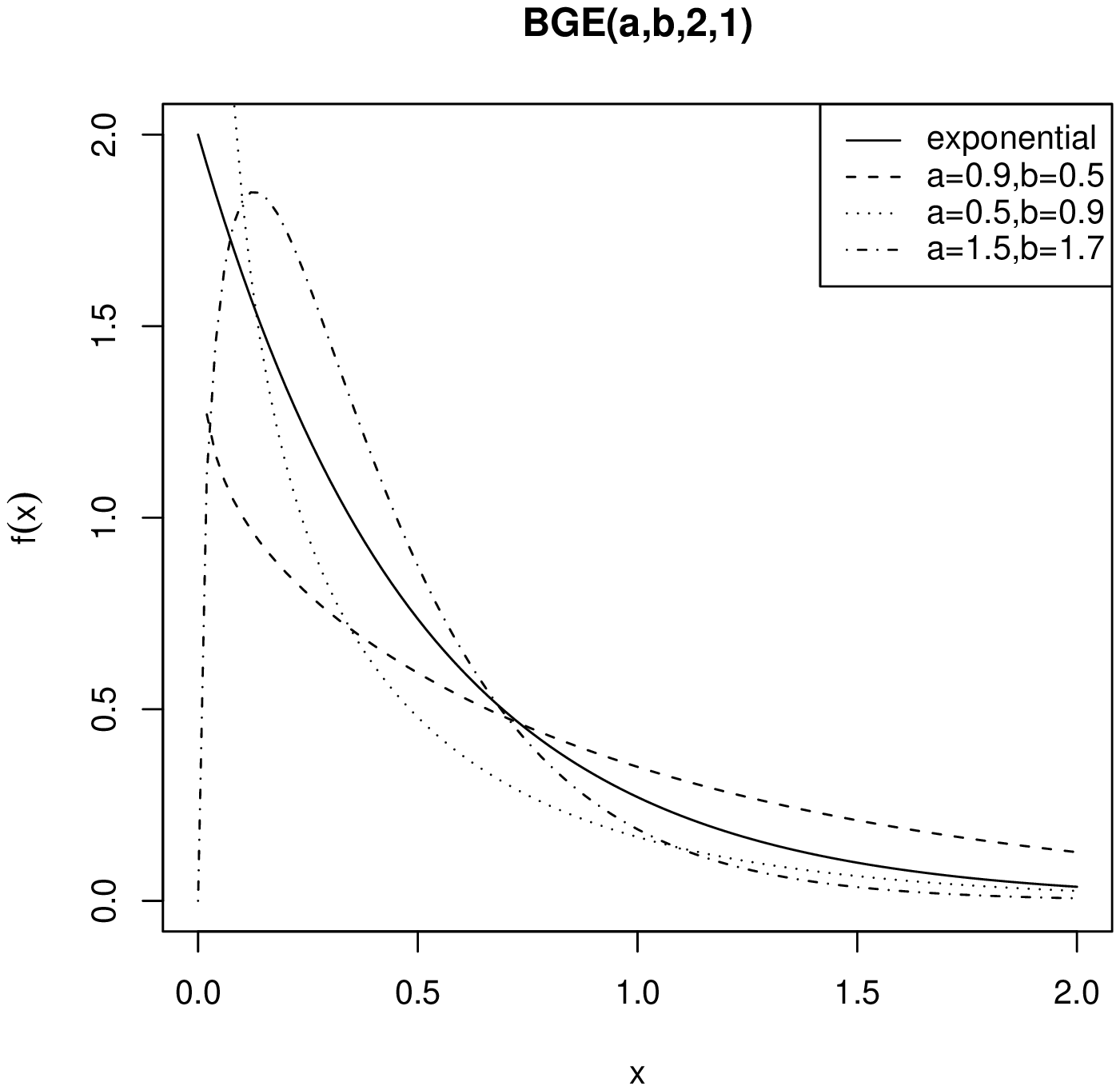}\includegraphics[scale=0.45]{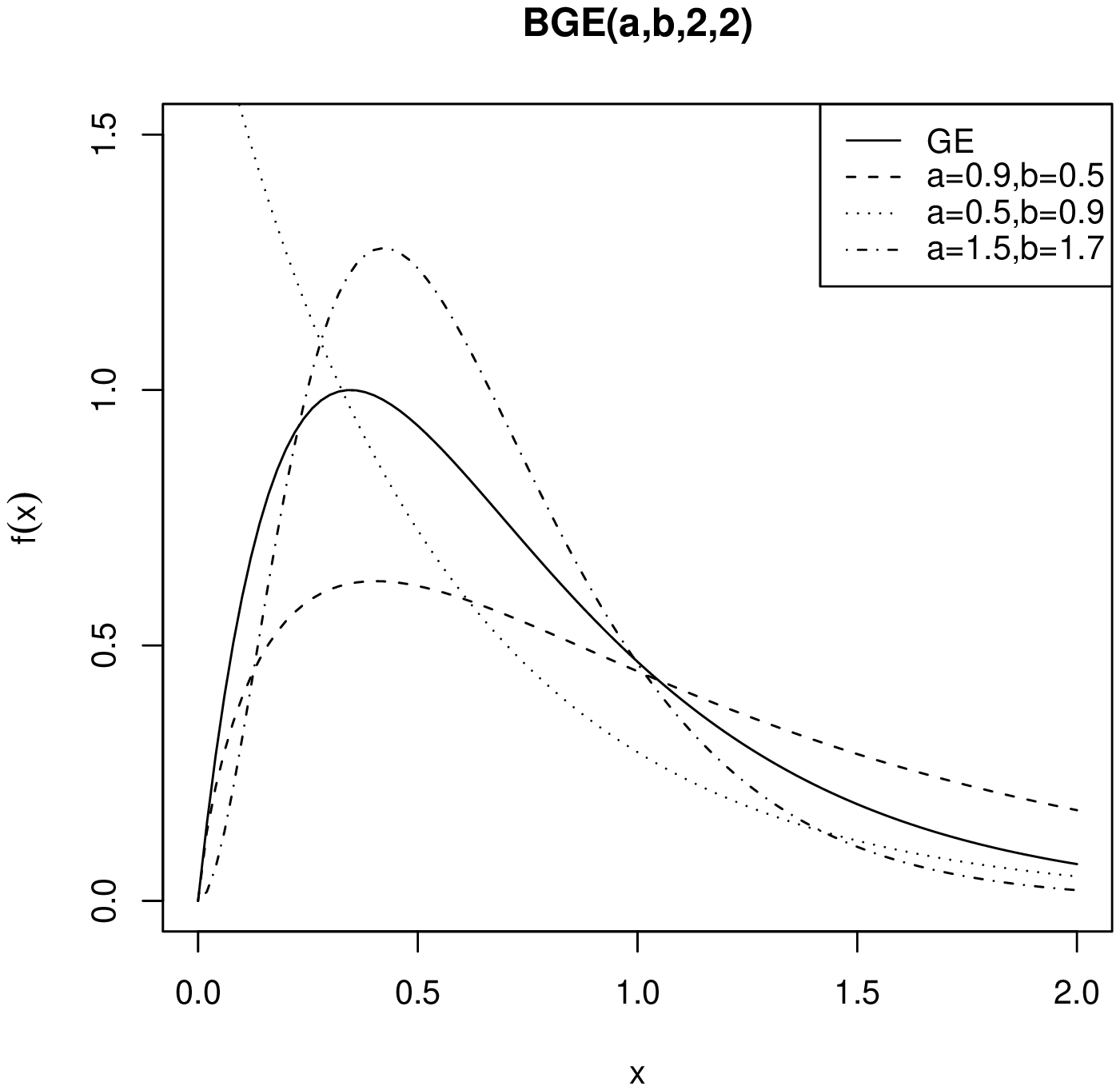}
\caption{Plots of the density (\ref{pdf_bee}) for some values of the parameters.}
\label{figpdf}
\end{figure}

\begin{figure}
\centering
\includegraphics[scale=0.45]{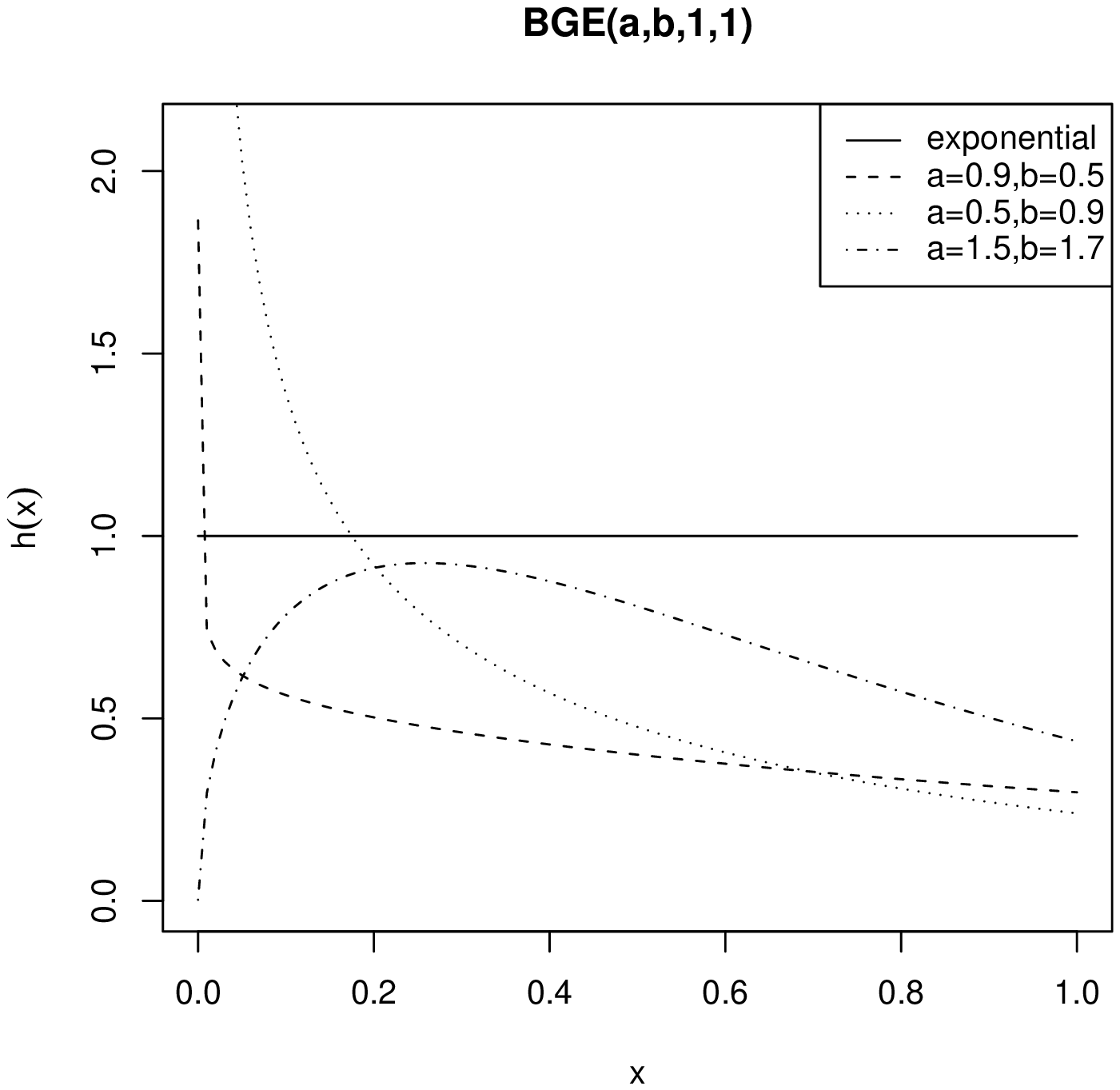}\includegraphics[scale=0.45]{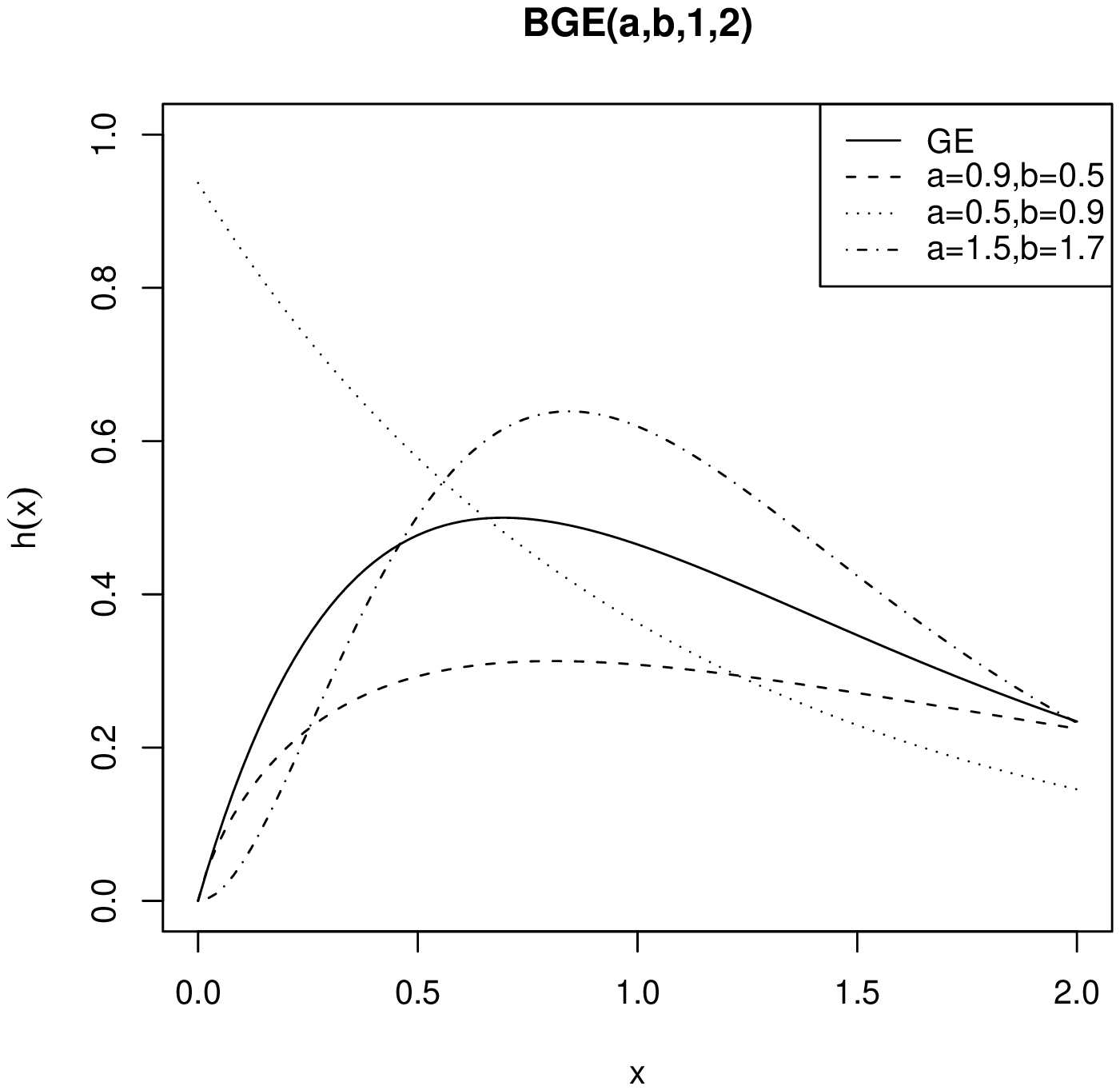}
\includegraphics[scale=0.45]{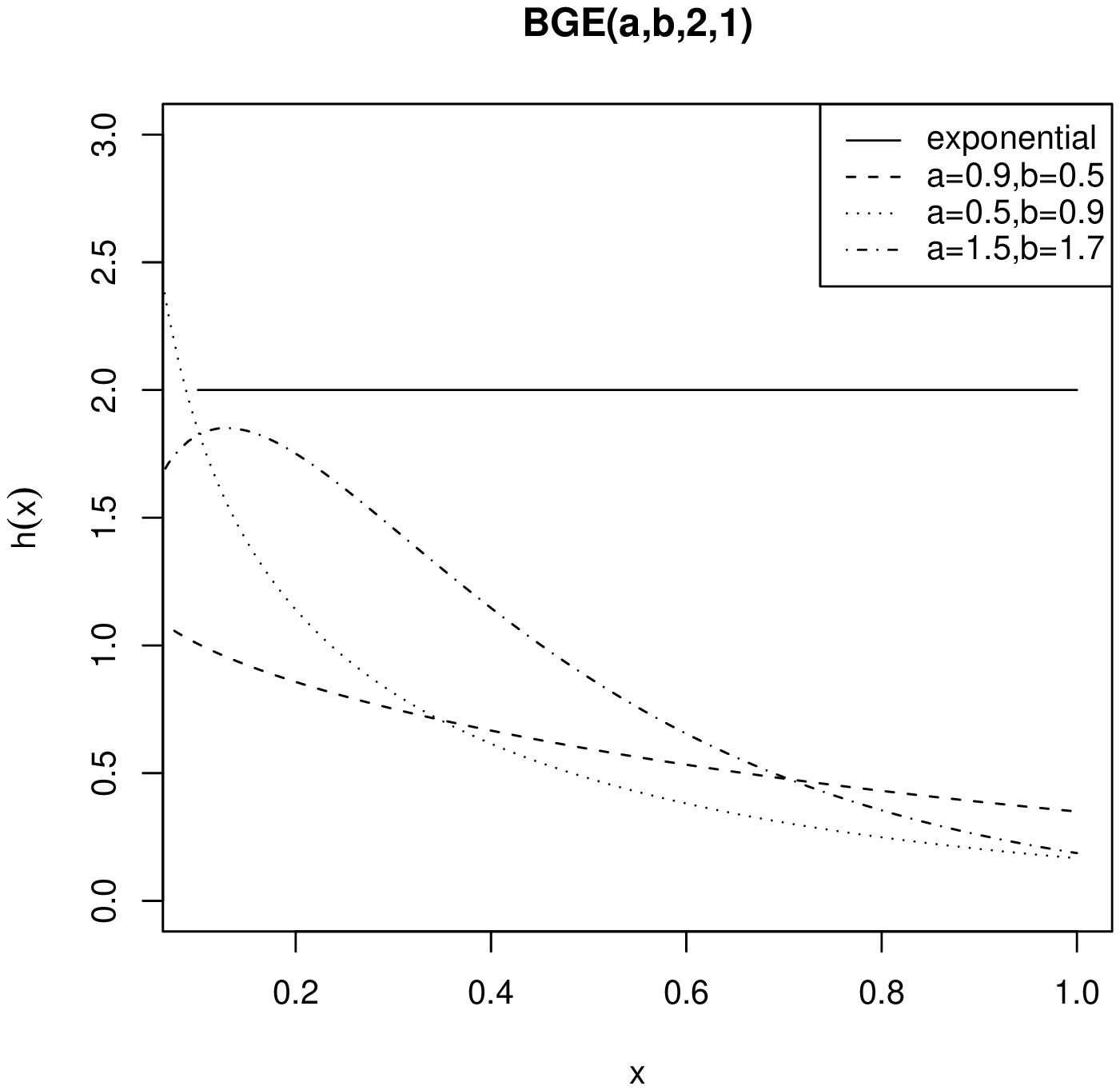}\includegraphics[scale=0.45]{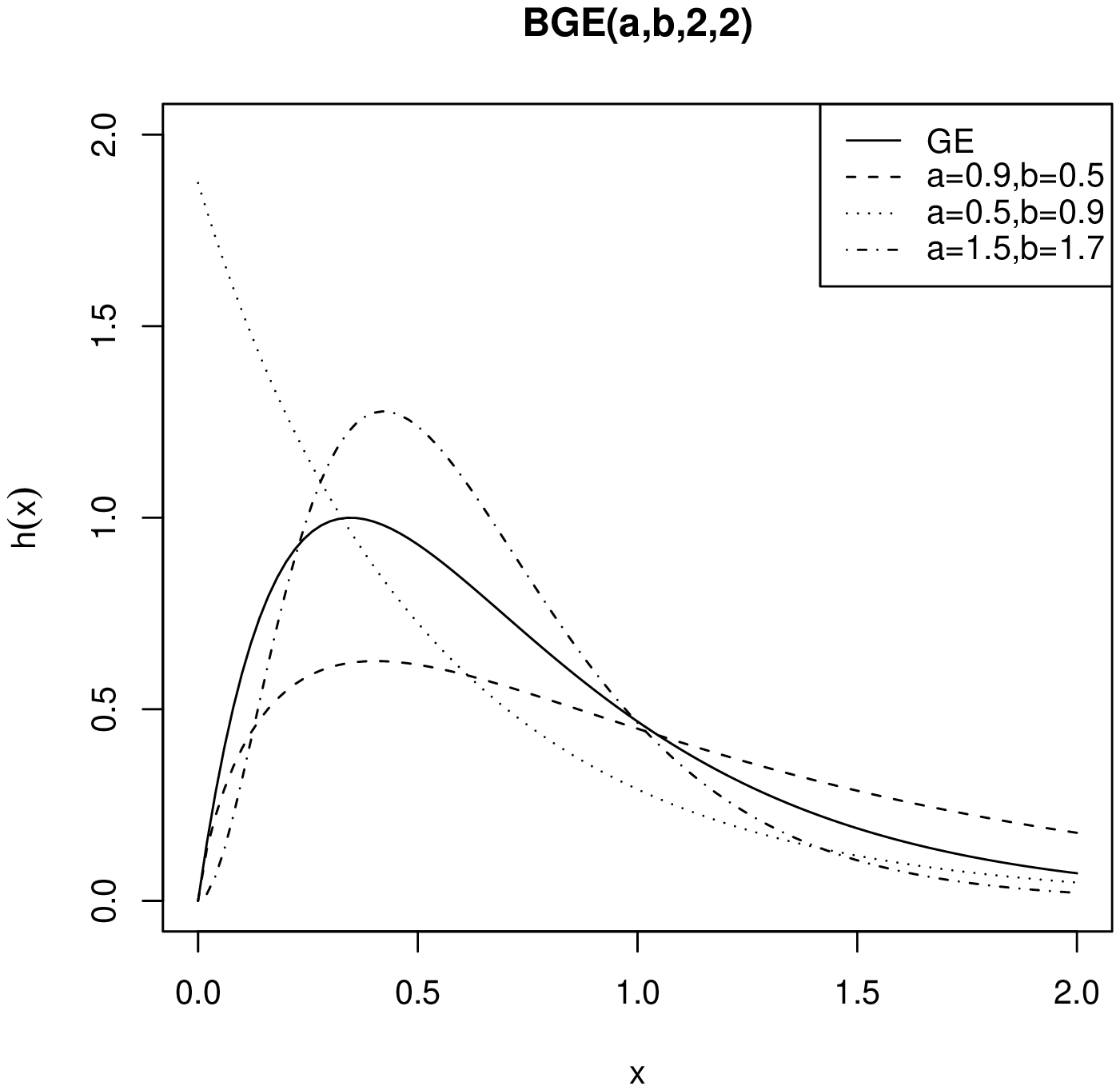}
\caption{Plots of the hazard (\ref{hazard_bee}) for some values of the parameters.}
\label{fighf}
\end{figure}

We now introduce the four parameter beta generalized exponential (BGE) distribution by
taking $G(x)$ in (\ref{beta_f}) to be the cdf (\ref{cd_ee}). The cdf of the
BGE distribution is then
\begin{equation}\label{cdf_bee}
F(x)=\frac{1}{B(a,b)}\int_0^{(1-e^{-\lambda x})^\alpha}\omega^{a-1}(1-\omega)^{b-1}d\omega,\quad x>0,
\end{equation}
for $a>0$, $b>0$, $\lambda>0$ and $\alpha>0$.
The pdf and the hazard rate function of the new distribution are, respectively,
\begin{equation}\label{pdf_bee}
f(x)=\frac{\alpha\lambda}{B(a,b)}e^{-\lambda x}(1-e^{-\lambda x})^{\alpha a-1}
\{1-(1-e^{-\lambda x})^\alpha \}^{b-1},\quad x>0,
\end{equation}
and
\begin{equation}\label{hazard_bee}
h(x)=\frac{\alpha\lambda e^{-\lambda x}(1-e^{-\lambda x})^{\alpha a-1}\left\{1-(1-e^{-\lambda x})^\alpha\right\}^{b-1}}{B(a,b)\, I_{1-(1-e^{-\lambda x})^\alpha}(a,b)},\quad x>0.
\end{equation}
The density function (\ref{pdf_bee}) does not involve any complicated function. If $X$ is a random
variable with pdf (\ref{pdf_bee}), we write $X\sim BGE(a,b,\lambda,\alpha)$. The BGE distribution 
generalizes some well-known distributions in the
literature. The GE distribution is a special case for the choice $a=b=1$. If in addition $\alpha=1$, we obtain the exponential
distribution with parameter $\lambda$. The BE distribution is obtained from
(\ref{cdf_bee}) with $\alpha=1$. It is evident that (\ref{pdf_bee}) is much more flexible than
the GE and BE distributions. Plots of the density
(\ref{pdf_bee}) and failure rate (\ref{hazard_bee}) for some special values of $a$, $b$, $\lambda$ and $\alpha$
are given in Figures \ref{figpdf} and \ref{fighf}, respectively. A characteristic of the BGE distribution is
that its failure rate can be bathtub shaped, monotonically increasing or
decreasing and upside-down bathtub depending basically on the values of its parameters.

We now introduce the double generalized exponential (DGE) distribution for the 
choice $a=1$ in (\ref{cdf_bee}). The cdf of the DGE distribution is
\begin{equation}\label{dee}
F(x)=\{1-(1-e^{-\lambda x})^\alpha \}^b,\quad x>0.
\end{equation}

The rest of the paper is organized as follows. In Section 2, we derive some expressions
for the cdf of the BGE distribution and for the pdfs of the order statistics.
In Sections 3 and 4 we provide expressions for the moment generating function and
for the $r$th moment, respectively. Furthermore, in these sections, we derive 
cor\-res\-pon\-ding expressions for the order statistics. In Section 5, we discuss maximum 
likelihood estimation, inference on the parameters, and calculate the information 
matrix. One application to real data set is presented in Section 6. Some conclusions are 
drawn in Section 7.

\section{Distribution function and order statistics}

We provide two simple formulae for the cdf of the BGE distribution depending if
the parameter $b>0$ is real non-integer or integer. First, if $|z|<1$ and $b>0$ is
real non-integer, we have
\begin{equation}\label{expansao}
(1-z)^{b-1}=\sum_{j=0}^\infty \frac{(-1)^j\Gamma(b)}{\Gamma(b-j)j!}z^j.
\end{equation}
Using the expansion (\ref{expansao}) in (\ref{cdf_bee}), the cdf of the BGE
distribution when $b>0$ is real non-integer follows
\begin{eqnarray*}
F(x)&=&\frac{1}{B(a,b)}\int_0^{(1-e^{-\lambda x})^\alpha}\omega^{a-1}(1-\omega)^{b-1}d\omega \nonumber\\
   &=&\frac{\Gamma(b)}{B(a,b)}\sum_{j=0}^\infty \frac{(-1)^j}{\Gamma(b-j)j!}\int_0^{(1-e^{-\lambda x})^\alpha}\omega^{a+j-1}d\omega\nonumber\\   
\end{eqnarray*}
and then
\vspace{-0.3cm}
\begin{eqnarray}
F(x)=\frac{\Gamma(a+b)}{\Gamma(a)}\sum_{j=0}^\infty \frac{(-1)^j(1-e^{-\lambda x})^{\alpha(a+j)}}{\Gamma(b-j)j!(a+j)}.
\label{cdfBGEreal}
\end{eqnarray}
Equation (\ref{cdfBGEreal}) reveals the property that the cdf of the BGE distribution can be
expressed as an infinite weighted sum of cdfs of GE distributions
\begin{eqnarray*}
F(x)=\frac{\Gamma(a+b)}{\Gamma(a)}\sum_{j=0}^{\infty}\frac{(-1)^jG_{\lambda,\alpha(a+j)}(x)}{\Gamma(b-j)j!(a+j)}.
\end{eqnarray*}
Using the binomial expansion in (\ref{cdf_bee}), we have when $b>0$ is integer
\begin{equation}\label{cdfBGEinteger}
F(x)=\frac{1}{B(a,b)}\sum_{j=0}^{b-1}\binom{b-1}{j}\frac{(-1)^j}{a+j}(1-e^{-\lambda x})^{\alpha(a+j)},
\end{equation}
and, again, the same property of equation (\ref{cdfBGEreal}) holds but now 
the sum is finite. Expressions (\ref{cdfBGEreal}) and (\ref{cdfBGEinteger}) 
are the main results of this section. The cdf of the BE distribution follows with $\alpha=1$ 
from (\ref{cdfBGEreal}) and (\ref{cdfBGEinteger}) depending if $b>0$ is real non-integer and 
integer, respectively. The cdf of the DGE distribution follows from the above expressions when $a=1$. 
The cdf (\ref{cd_ee}) of the GE distribution comes from (\ref{cdfBGEinteger}) when $a=b=1$.
If, in addition, $\alpha=1$, (\ref{cdfBGEinteger}) yields the cdf $F(x)=1-e^{-\lambda x}$ of 
the exponential distribution.

It can be seen in the Wolfram Functions Site\footnote{{\tt http://functions.wolfram.com/}} that for integer $a$
\begin{equation*}
I_y(a,b)=1-\frac{(1-y)^b}{\Gamma(b)}\sum_{j=0}^{a-1}\frac{\Gamma(b+j)}{j!}y^j,
\end{equation*}
and for integer $b$
\begin{equation*}
I_y(a,b)=\frac{y^a}{\Gamma(a)}\sum_{j=0}^{b-1}\frac{\Gamma(a+j)}{j!}(1-y)^j.
\end{equation*}
Therefore, for $a>0$ integer,
\begin{equation*}
F(x)=1-\frac{\left\{1-(1-e^{-\lambda \,x})^\alpha\right\}^b}{\Gamma(b)}\sum_{j=0}^{a-1}\frac{\Gamma(b+j)}{j!}(1-e^{-\lambda x})^{\alpha j}
\end{equation*}
and for $b>0$ integer we have an alternative form for (\ref{cdfBGEinteger}) given by
\begin{equation*}
F(x)=\frac{(1-e^{-\lambda x})^{a\alpha}}{\Gamma(a)}\sum_{j=0}^{b-1}\frac{\Gamma(a+j)}{j!}\{1-(1-e^{-\lambda x})^\alpha\}^j.
\end{equation*}

The density function (\ref{pdf_bee}) can be expressed in the mixture form in terms of 
cdfs of the GE distributions
\begin{eqnarray*}
f(x)=\frac{\alpha\lambda}{B(a,b)}e^{-\lambda x}G_{\lambda,a\alpha-1}(x)\sum_{j=0}^{\infty}\frac{(-1)^j\Gamma(b)}{\Gamma(b-j)j!}G_{\lambda,\alpha j}(x)
\end{eqnarray*}
for $b$ real non-integer and
\begin{eqnarray*}
f(x)=\frac{\alpha\lambda}{B(a,b)}e^{-\lambda x}G_{\lambda,a\alpha-1}(x)\sum_{j=0}^{b-1}\binom{b-1}{j}(-1)^jG_{\lambda,\alpha j}(x)
\end{eqnarray*}
for $b$ integer.\\

Simulation of the BGE distribution is easy: if $V$ is a random variable with a beta distribution
with parameters $a$ and $b$, then $X=-\log(1-V^{1/\alpha})/\lambda$ follows the BGE distribution
with parameters $a$, $b$, $\lambda$ and $\alpha$.

We now give the density of the $i$th order statistic $X_{i:n}$, $f_{i:n}(x)$ say, 
in a random sample of size $n$ from the BGE distribution. It is well known that
\begin{eqnarray*}
f_{i:n}(x)=\frac{1}{B(i,n-i+1)}f(x)F^{i-1}(x)\left\{1-F(x)\right\}^{n-i},
\end{eqnarray*}
for $i=1,\ldots,n$. Using (\ref{cdf_*A}) and (\ref{pdf_bee}) we can express $f_{i:n}(x)$
in terms of the hyper\-geo\-me\-tric functions by
\begin{eqnarray*}
f_{i:n}(x)&=&\frac{\alpha\lambda e^{-\lambda x}G_{\lambda,a\alpha-1}(x)G_{\lambda,\alpha a(i-1)}(x)\{1-G_{\lambda,\alpha}(x)\}^{b(n-i+1)-1}}{B(i,n-i+1)B(a,b)^na^{i-1}b^{n-i}}\times\\
&&_2F_1(a,1-b,a+1;G_{\lambda,\alpha}(x))^{i-1}\,_2F_1(b,1-a,b+1;1-G_{\lambda,\alpha}(x))^{n-i}.
\end{eqnarray*}

We now derive two alternative expressions for the densities of the order statistics using 
the expansion $(\sum_{i=1}^\infty a_i)^k=\sum_{\{m_1,\ldots,m_k\}=0}^\infty a_{m_1}\ldots a_{m_k}$ 
for $k$ a positive integer. With this expansion and from (\ref{cdfBGEreal})$-($\ref{cdfBGEinteger}), 
we can show for $b>0$ real non-integer and integer that
\begin{eqnarray}\label{pdforderreal}
f_{i:n}(x)=\sum_{k=0}^{n-i}\sum_{m_1=0}^{\infty}\ldots\sum_{m_{k+i-1}=0}^{\infty}\delta^{(1)}_{k,i} f_{k,i}(x)
\end{eqnarray}
and
\begin{eqnarray}\label{pdforderinteger}
f_{i:n}(x)=\sum_{k=0}^{n-i}\sum_{m_1=0}^{b-1}\ldots\sum_{m_{k+i-1}=0}^{b-1}\delta^{(2)}_{k,i} f_{k,i}(x),
\end{eqnarray}
respectively, where from now on $f_{k,i}(x)$ represents the density of a random variable $X_{k,i}$ 
following a $BGE(\alpha\{a(i+1)+\sum_{j=1}^{k+i-1}m_j\},b,\lambda,\alpha)$ distribution, and the 
functions 
$\delta^{(1)}_{k,i}$ and $\delta^{(2)}_{k,i}$ required for the above expressions are $$\delta^{(1)}_{k,i}=\frac{(-1)^{k+\sum_{j=1}^{k+i-1}m_j}\binom{n-i}{k}B(\alpha\{a(i+1)+\sum_{j=1}^{k+i-1}m_j\},b)\Gamma(b)^{k+i-1}}{B(a,b)^{k+i}B(i,n-i+1)\prod_{j=1}^{k+i-1}\Gamma(b-m_j)m_j!(a+m_j)}$$ 
and 
$$\delta^{(2)}_{k,i}=\frac{(-1)^{k+\sum_{j=1}^{k+i-1}m_j}\binom{n-i}{k}B(\alpha\{a(i+1)+\sum_{j=1}^{k+i-1}m_j\},b)}{B(a,b)^{k+i}B(i,n-i+1)}\prod_{j=1}^{k+j-1}\frac{\binom{b-1}{m_j}}{(a+m_j)}.$$ 
The sums in (\ref{pdforderreal}) and (\ref{pdforderinteger}) extends over all 
$(k+i)$-tuples ($k, m_1,\ldots,m_{k+i-1}$) of non-negative integers and are 
easily implementable on a computer. 

\section{Moment generating function}

The moment generating function (mgf) of the BGE distribution is given by
\begin{eqnarray}\label{mgf}
M(t)=\frac{\alpha\lambda}{B(a,b)}\int_0^\infty e^{tx}e^{-\lambda x}(1-e^{-\lambda x})^{\alpha a-1}\{1-(1-e^{-\lambda x})^\alpha\}^{b-1}dx.
\end{eqnarray}
Using the expansion $(\ref{expansao})$ when $b>0$ is real
non-integer, (\ref{mgf}) reduces to
\begin{eqnarray*}
M(t)=\frac{\alpha\lambda\Gamma(b)}{B(a,b)}\sum_{j=0}^\infty \frac{(-1)^j}{\Gamma(b-j)j!}\int_0^\infty e^{(t-\lambda)x}(1-e^{-\lambda x})^{\alpha j}dx.
\end{eqnarray*}
Changing to the variable $u=e^{-\lambda x}$, we obtain
\begin{eqnarray*}
M(t)=\frac{\alpha\Gamma(b)}{B(a,b)}\sum_{j=0}^\infty \frac{(-1)^j}{\Gamma(b-j)j!}\int_0^1 u^{-t/\lambda}(1-u)^{\alpha(a+j)-1}du.
\end{eqnarray*}
The above expression shows that the mgf of the BGE distribution exists if $t<\lambda$. Assuming that
$t<\lambda$, we have
\begin{equation}\label{mgfreal}
M(t)=\frac{\alpha\Gamma(b)}{B(a,b)}\sum_{j=0}^\infty \frac{(-1)^j}{\Gamma(b-j)j!}B(1-t/\lambda,\alpha(a+j)).
\end{equation}
Analogously, if $b>0$ is integer, assuming $t<\lambda$, we use the binomial expansion in (\ref{mgf}) to obtain
\begin{equation}\label{mgfinteger}
M(t)=\frac{\alpha}{B(a,b)}\sum_{j=0}^{b-1}\binom{b-1}{j}(-1)^jB(1-t/\lambda,\alpha(a+j)).
\end{equation}
If we take $a=b=1$ in (\ref{mgfinteger}), the above mgf reduces to
\begin{equation*}
M(t)=\alpha B(1-t/\lambda,\alpha),
\end{equation*}
which agrees with Gupta and Kundu's \cite{Gupta2001} equation (2.3).

From (\ref{mgfreal}) and (\ref{mgfinteger}) with $\alpha=1$ we have the mgf of the BE distribution
\begin{eqnarray}\label{mgfbe}
M(t)&=&\frac{\Gamma(b)}{B(a,b)}\sum_{j=0}^\infty\frac{(-1)^j}{\Gamma(b-j)j!}B(1-t/\lambda,a+j)\nonumber\\
&=&\frac{B(b-t/\lambda,a)}{B(a,b)},
\end{eqnarray}
which agrees with Nadarajah and Kotz's \cite{Nadarajah2005} equation (3.1). The last equality 
is given with more details in the Appendix A. The expressions for the mgf of the DGE distribution are 
obtained with $a=1$ in (\ref{mgfreal}) and (\ref{mgfinteger}) for $b>0$ real non-integer and $b>0$ integer, 
respectively.

We can easily obtain expressions for the mgf of the order statistics from (\ref{pdforderreal}) 
and (\ref{pdforderinteger}). The mgf of $X_{i:n}$ for $b>0$ real non-integer is 
\begin{eqnarray*}
M_{X_{i:n}}(t)=\sum_{k=0}^{n-i}\sum_{m_1=0}^{\infty}\ldots\sum_{m_{k+i-1}=0}^{\infty}\delta^{(1)}_{k,i} M_{k,i}(t)
\end{eqnarray*}
and for $b>0$ integer
\begin{eqnarray*}
M_{X_{i:n}}(t)=\sum_{k=0}^{n-i}\sum_{m_1=0}^{b-1}\ldots\sum_{m_{k+i-1}=0}^{b-1}\delta^{(2)}_{k,i}M_{k,i}(t),
\end{eqnarray*} 
where $M_{k,i}(t)$ is the mgf of the $BGE(\alpha\{a(i+1)+\sum_{j=1}^{k+i-1}m_j\},b,\lambda,\alpha)$ distribution. 

\section{Moments}

The $r$th moment of the BGE distribution can be obtained from $E(X^r)=d^rM_X(t)/dt^r|_{t=0}$.
Hence, if $b>0$ is real non-integer, we have from (\ref{mgfreal})
\begin{equation}\label{momentreal}
\mu_r'=E(X^r)=\frac{\alpha\Gamma(b)}{\lambda^rB(a,b)}\sum_{j=0}^\infty\frac{(-1)^{j+r}}{\Gamma(b-j)j!}\frac{d^rB(p,\alpha(a+j))}{dp^r}\bigg|_{p=1}
\end{equation}
and if $b>0$ is integer, we obtain from (\ref{mgfinteger})
\begin{equation}\label{momentinteger}
\mu_r'=E(X^r)=\frac{\alpha\Gamma(b)}{\lambda^rB(a,b)}\sum_{j=0}^{b-1}\binom{b-1}{j}(-1)^{j+r}\frac{d^rB(p,\alpha(a+j))}{dp^r}\bigg|_{p=1}.
\end{equation}
Equations (\ref{momentreal}) and (\ref{momentinteger}) are the main results of this
section and generalize the moments of the GE and BE distributions derived by Gupta and Kundu \cite{Gupta2001} and Nadarajah and Kotz \cite{Nadarajah2005}, respectively. The first four moments of
the BGE distribution for $b>0$ real non-integer are
\begin{eqnarray*}
&&\mu_1'=\frac{\Gamma(a+b)}{\lambda\Gamma(a)}\sum_{j=0}^\infty \frac{(-1)^j(a+j)^{-1}}{\Gamma(b-j)j!}c_j,\quad \mu_2'=\frac{\Gamma(a+b)}{\lambda^2\Gamma(a)}\sum_{j=0}^\infty \frac{(-1)^j(a+j)^{-1}}{\Gamma(b-j)j!}d_j,\\
&&\mu_3'=\frac{\Gamma(a+b)}{\lambda^3\Gamma(a)}\sum_{j=0}^\infty \frac{(-1)^{j+1}(a+j)^{-1}}{\Gamma(b-j)j!}e_j,
\quad\mu_4'=\frac{\Gamma(a+b)}{\lambda^4\Gamma(a)}\sum_{j=0}^\infty \frac{(-1)^j(a+j)^{-1}}{\Gamma(b-j)j!}f_j,
\end{eqnarray*}
and for $b>0$ integer are
\begin{eqnarray*}
&&\mu_1'=\frac{\Gamma(a+b)}{\lambda\Gamma(a)}\sum_{j=0}^{b-1}\binom{b-1}{j}\frac{(-1)^j}{a+j}c_j, \quad\mu_2'=\frac{\Gamma(a+b)}{\lambda^2\Gamma(a)}\sum_{j=0}^{b-1}\binom{b-1}{j}\frac{(-1)^j}{a+j}d_j,\\
&&\mu_3'=\frac{\Gamma(a+b)}{\lambda^3\Gamma(a)}\sum_{j=0}^{b-1}\binom{b-1}{j}\frac{(-1)^{j+1}}{a+j}e_j,
\quad\mu_4'=\frac{\Gamma(a+b)}{\lambda^4\Gamma(a)}\sum_{j=0}^{b-1}\binom{b-1}{j}\frac{(-1)^j}{a+j}f_j,
\end{eqnarray*}
where the quantities $c_j$, $d_j$, $e_j$ and $f_j$ are given in the Appendix B.

\begin{figure}
\centering
\includegraphics[width=0.45\textwidth]{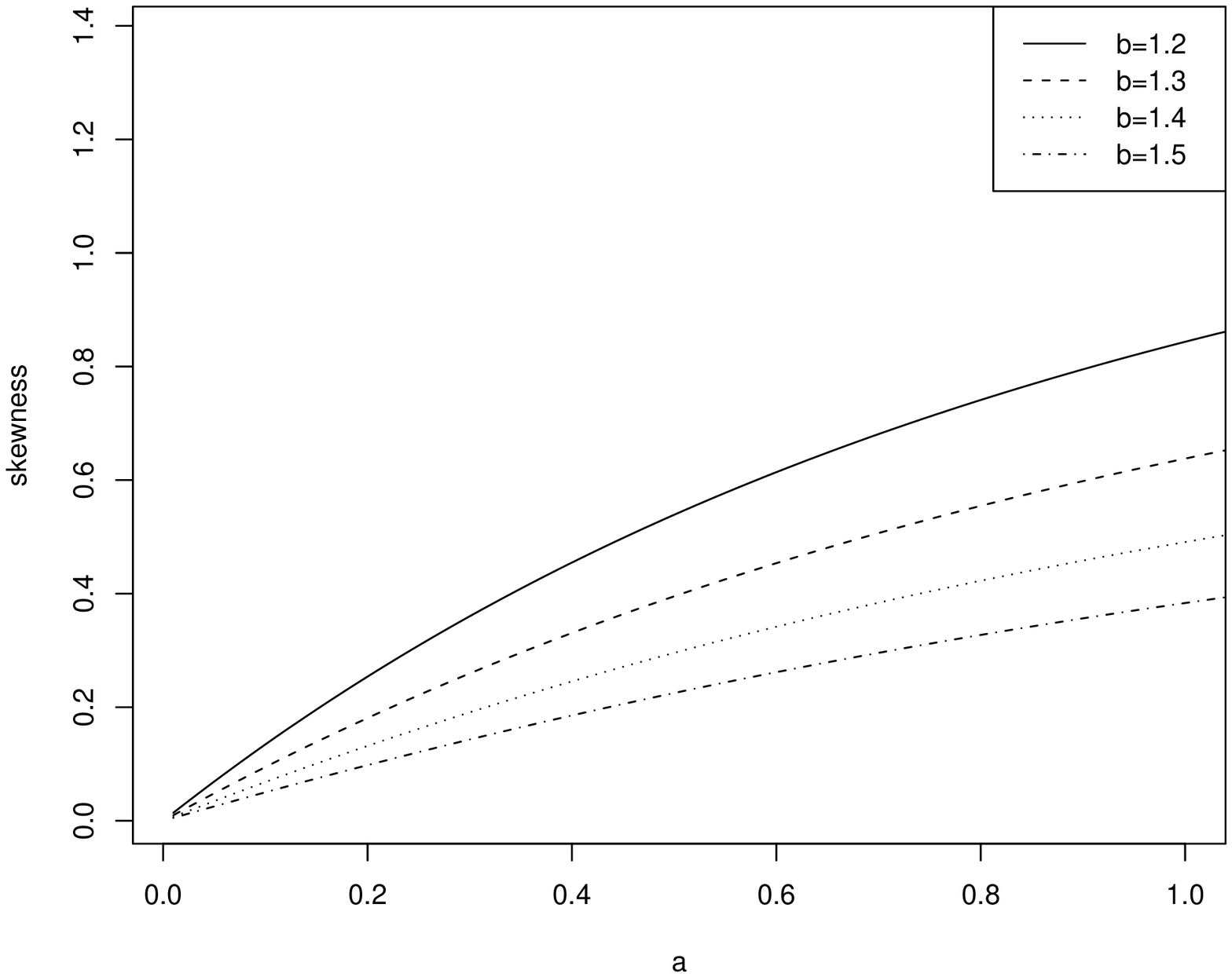}\includegraphics[width=0.45\textwidth]{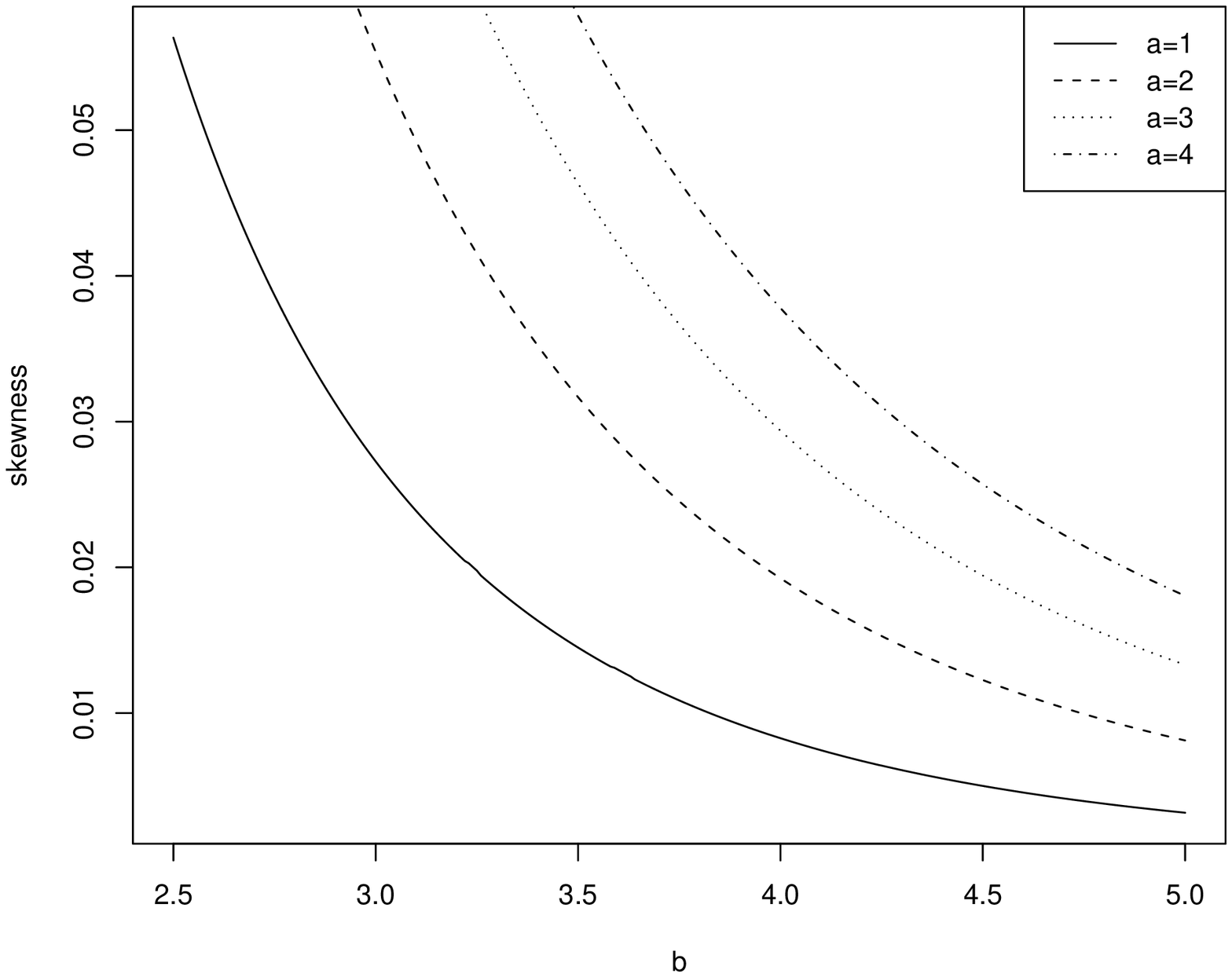}
\caption{Skewness of the BGE distribution as function of $a$ for some values of $b$ and as function of $b$ for some values of $a$.}
\label{figskewness}
\end{figure}

\begin{figure}
\centering
\includegraphics[width=0.45\textwidth]{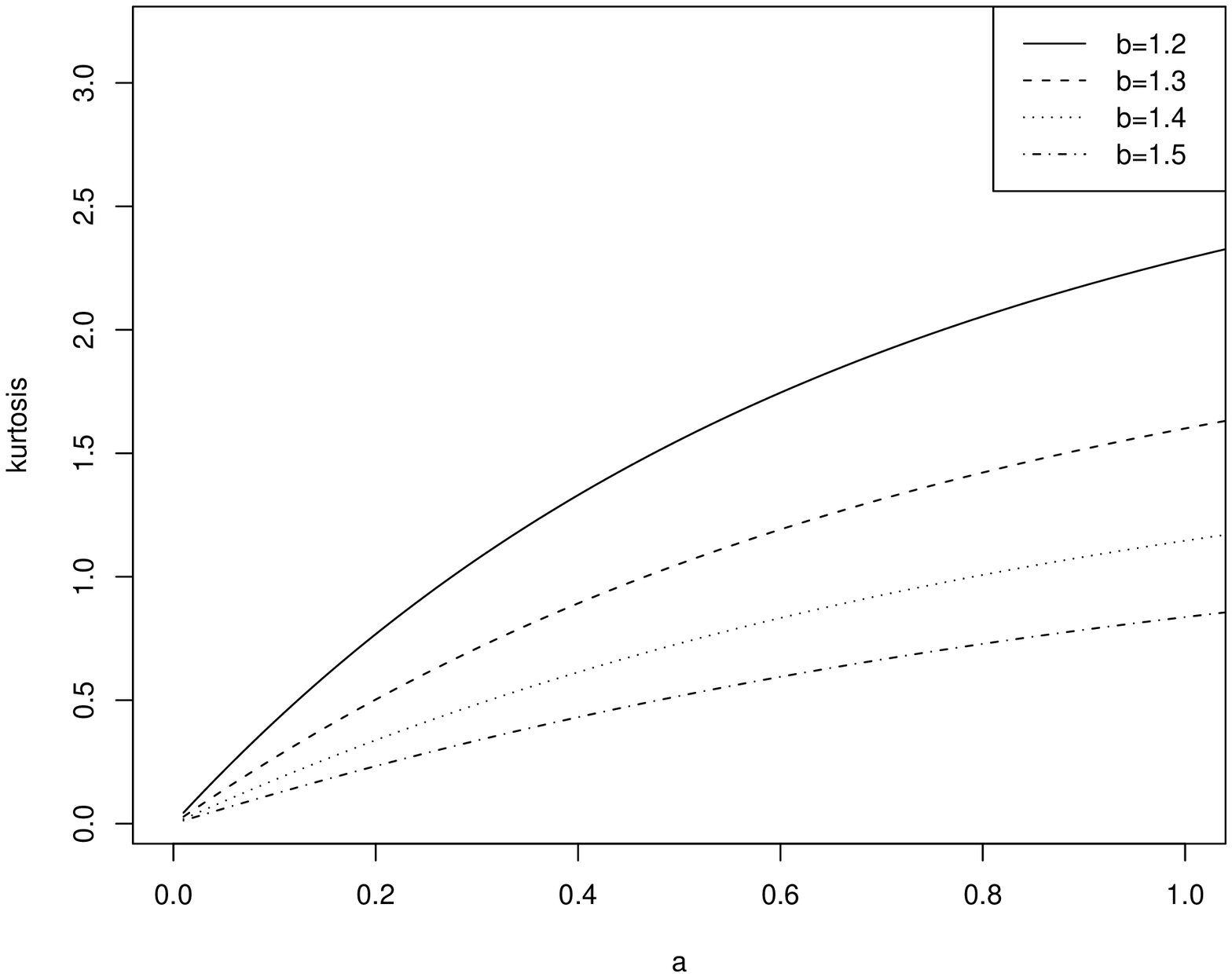}\includegraphics[width=0.45\textwidth]{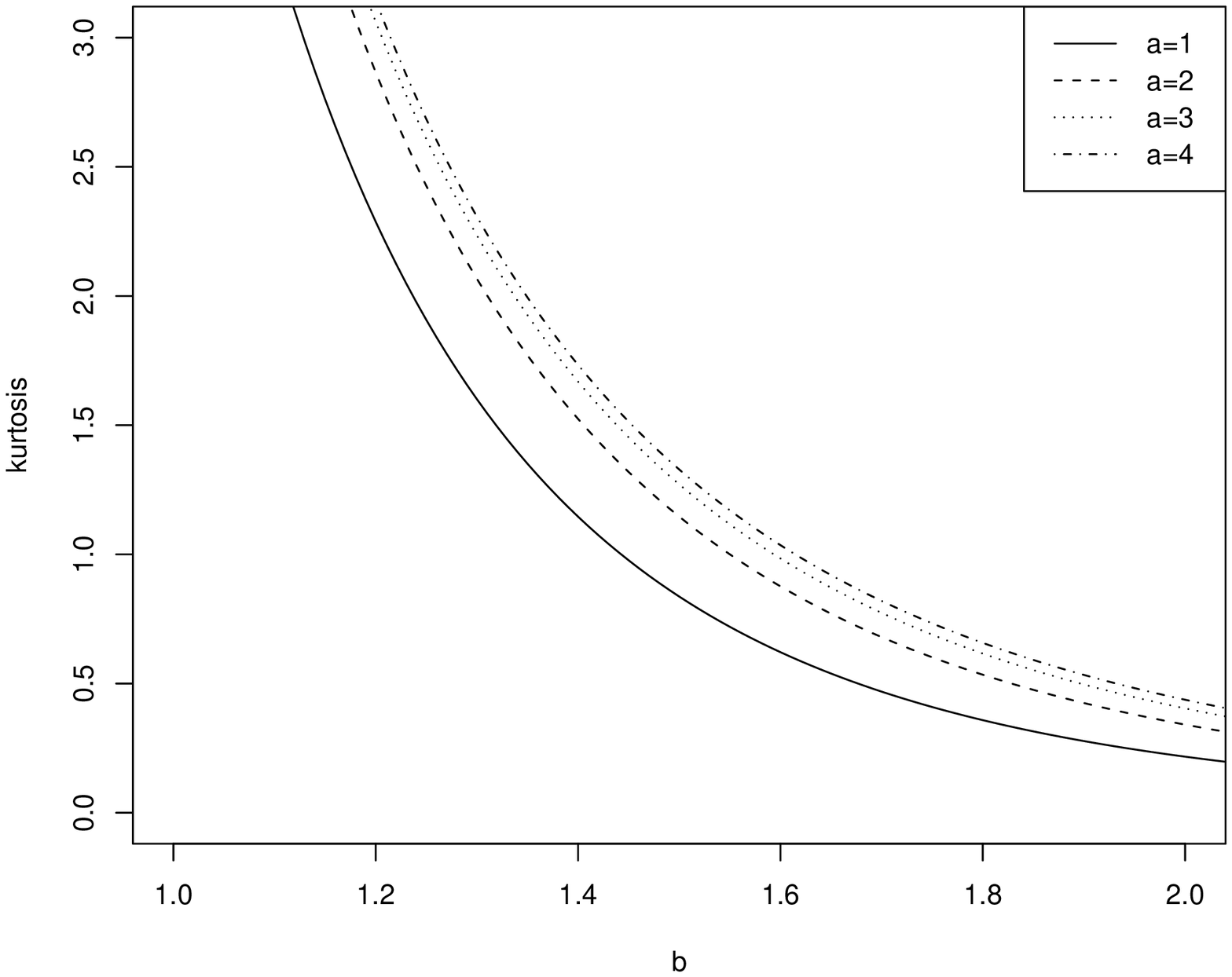}
\caption{Kurtosis of the BGE distribution as function of $a$ for some values of $b$ and as function of $b$ for some values of $a$.}
 \label{figkurtosis}
\end{figure}

Graphical representation of skewness and kurtosis for some choices of 
parameter $b$ as function of parameter $a$, and for some choices of parameter $a$ as function 
of parameter $b$, for fixed $\lambda=1$ and $\alpha=1$, are given in Figures \ref{figskewness} 
and \ref{figkurtosis}, respectively. These plots show that the skewness and kurtosis curves 
increase (decrease) with $a$ ($b$) for fixed $b$ ($a$).

The $r$th moment of $X_{i:n}$ for $b>0$ real non-integer is 
\begin{eqnarray*}
E(X_{i:n}^r)=\sum_{k=0}^{n-i}\sum_{m_1=0}^{\infty}\ldots\sum_{m_{k+i-1}=0}^{\infty}\delta^{(1)}_{k,i}E(X_{k,i}^r)
\end{eqnarray*}
and for $b>0$ integer
\begin{eqnarray*}
E(X_{i:n}^r)=\sum_{k=0}^{n-i}\sum_{m_1=0}^{b-1}\ldots\sum_{m_{k+i-1}=0}^{b-1}\delta^{(2)}_{k,i}E(X_{k,i}^r),
\end{eqnarray*} 
where the moments $E(X_{k,i}^r)$ come from the general expansions (\ref{momentreal}) and (\ref{momentinteger}) 
for the moments of the BGE distribution with parameters $\alpha\{a(i+1)+\sum_{j=1}^{k+i-1}m_j\},b,\lambda$ and $\alpha$. \\

The Shannon entropy of a random variable $X$ is a measure of the uncertainty and 
is defined by $E\{-\log f(X)\}$, where $f(x)$ is the fdp of the $X$. For a random 
variable $X$ with a BGE distribution, we obtain
\begin{eqnarray*}
E\{-\log f(X)\}&=&-\log(\alpha\lambda)+\log B(a,b)+\lambda\mu_1'+(1/\alpha-a)\left\{\Psi(a)-\Psi(a+b)\right\}\\
&&-(b-1)\left\{\Psi(b)-\Psi(a+b)\right\},
\end{eqnarray*}
where $\mu_1'$ is the mean given before.

\section{Estimation and Inference}

We assume that $Y$ follows the BGE distribution and let $\theta=(a,b,\lambda,\alpha)^T$ be the
parameter vector. The log-likelihood for a single observation $y$ of $Y$ is
\begin{eqnarray*}
\ell=\ell(a,b,\lambda,\alpha)&=&\log\alpha+\log\lambda-\log B(a,b)-\lambda y+(\alpha a-1)\log(1-e^{-\lambda y})\nonumber\\
&&+(b-1)\log\{1-(1-e^{-\lambda y})^\alpha\},\quad y>0.
\end{eqnarray*}
The components of the unit score vector $U=(\frac{\partial \ell}{\partial a},\frac{\partial \ell}{\partial b},\frac{\partial \ell}{\partial \lambda},\frac{\partial \ell}{\partial \alpha})^T$ are
\begin{eqnarray*}
\frac{\partial \ell}{\partial a} &=& \alpha \log(1-e^{-\lambda y}) - \psi(a) + \psi(a+b),\\
\frac{\partial \ell}{\partial b} &=& \log\{1-(1-e^{-\lambda y})^{\alpha}\} - \psi(b) + \psi(a+b),\\
\frac{\partial \ell}{\partial \lambda} &=& \frac{1}{\lambda} - y + \frac{(\alpha a - 1) y e^{-\lambda y}}{1-e^{-\lambda y}} - \frac{\alpha (b-1)e^{-\lambda y}(1-e^{-\lambda y})^{\alpha-1}y}{1-(1-e^{-\lambda y})^{\alpha}},\\
\frac{\partial \ell}{\partial \alpha} &=& \frac{1}{\alpha} + a \log(1-e^{-\lambda y}) - \frac{(b-1)(1-e^{-\lambda y})^{\alpha} \log(1-e^{-\lambda y})}{1-(1-e^{-\lambda y})^{\alpha}}.
\end{eqnarray*}
The expected value of the score vector vanishes and then
\begin{eqnarray*}
E\{\log(1-e^{-\lambda Y})\}&=&\frac{\psi(a)-\psi(a+b)}{\alpha},\\
E[\log\{1-(1-e^{-\lambda Y})^\alpha\}]&=&\psi(b)-\psi(a+b),\\
E\bigg\{\frac{(1-e^{-\lambda Y})^\alpha\log(1-e^{-\lambda Y})}{1-(1-e^{-\lambda Y})^\alpha}\bigg\}&=&\frac{a\left\{\psi(a)-\psi(a+b)\right\}+1}{\alpha(b-1)}.
\end{eqnarray*}
For a random sample $y=(y_1,\ldots,y_n)$ of size $n$ from $Y$, the total log-likelihood is
$\ell_n=\ell_n(a,b,\lambda,\alpha)=\sum_{i=1}^n \ell^{(i)},$
where $\ell^{(i)}$ is the log-likelihood for the $i$th observation ($i=1,\ldots,n$). The total
score function is $U_n=\sum_{i=1}^nU^{(i)}$, where $U^{(i)}$ has the form given before for
$i=1,\ldots,n$. The MLE $\hat\theta$ of $\theta$ is obtained numerically from the nonlinear
equations $U_n=0$. For interval estimation and tests of hypotheses
on the parameters in $\theta$ we obtain the $4\times4$ unit information matrix
\begin{eqnarray*}\label{fisher}
K = K(\theta) = \left[ \begin{array}{cccc}
\kappa_{a,a}&\kappa_{a,b}&\kappa_{a,\lambda}&\kappa_{a,\alpha}\\
\kappa_{a,b}&\kappa_{b,b}&\kappa_{b,\lambda}&\kappa_{b,\alpha}\\
\kappa_{a,\lambda}&\kappa_{b,\lambda}&\kappa_{\lambda,\lambda}&\kappa_{\lambda,\alpha}\\
\kappa_{a,\alpha}&\kappa_{b,\alpha}&\kappa_{\lambda,\alpha}&\kappa_{\alpha,\alpha}
\end{array}\right],
\end{eqnarray*}
where the corresponding elements are given by
\begin{eqnarray*}
\kappa_{a,a}&=&\psi'(a)-\psi'(a+b),\quad \kappa_{a,b}=-\psi'(a+b), \quad\kappa_{a,\lambda}=\frac{\alpha}{\lambda}T_{0,1,1,1,0}\\
\kappa_{a,\alpha}&=&\frac{\psi(a+b)-\psi(a)}{\alpha},\quad\kappa_{b,b}=\psi'(b)-\psi'(a+b),\quad\kappa_{b,\lambda}=-\frac{\alpha}{\lambda}T_{1,1,1,1,0,}\\
\kappa_{\lambda,\lambda}&=&\frac{1}{\lambda^2}[1+(\alpha a-1)(T_{0,2,2,2,0}+T_{0,1,1,2,0})+\alpha(b-1)\{\alpha T_{2,2,2,2,0}+\\&&(\alpha-1)T_{1,2,2,2,0}-T_{1,1,1,2,0}\}],\quad\kappa_{b,\alpha}=\frac{a\left\{\psi(a)-\psi(a+b)\right\}+1}{\alpha(b-1)},\\
\kappa_{\lambda,\alpha}&=&\frac{1}{\lambda}\{aT_{0,1,1,1,0}-(b-1)(T_{1,1,1,1,0}+T_{2,1,1,1,1}+T_{1,1,1,1,1})\},\\
\kappa_{\alpha,\alpha}&=&\frac{1}{\alpha^2}\{1+(b-1)(T_{2,0,0,0,2}+T_{1,0,0,0,2})\}.
\end{eqnarray*}
Here, we have defined the following expectation
$$T_{i,j,k,l,m}=E[(1-V)^{-i} (1-V^{1/\alpha})^j V^{i-k/\alpha}\{\log(1-V^{1/\alpha})\}^l(\log V)^m],$$
where $V\sim Beta(a,b)$ and $i,j,k,l,m\in\{0,1,2\}$. The total information matrix is
then $K_n =K_n(\theta)=n K(\theta)$.

Under conditions that are fulfilled for parameters in
the interior of the parameter space but not on
the boundary, the asymptotic distribution of
$$ \sqrt n (\hat\theta-\theta)\,\,\,\,\mathrm{is}\,\,\,\,N_4(0,K(\theta)^{-1}).$$
The asymptotic multivariate normal $N_4(0,K_n(\hat\theta)^{-1})$
distribution of $\hat\theta$ can be used to construct approximate
confidence intervals and confidence regions for the parameters and
for the hazard and survival functions. The asymptotic normality is
also useful for testing goodness of fit of the BGE
distribution and for comparing this distribution with some of its
special sub-models using one of the three well-known asymptotically
equivalent test statistics - namely, the likelihood ratio (LR)
statistic, Rao score ($S_R$) and Wald ($W$) statistics.

An asymptotic confidence interval with significance level $\gamma$ for each
parameter $\theta_i$ is given by
$$ACI(\theta_i,100(1-\gamma)\%)=(\hat{\theta_i}-z_{\gamma/2}\sqrt{\kappa^{\theta_i,\theta_i}},\hat{\theta_i}+z_{\gamma/2}\sqrt{\kappa^{\theta_i,\theta_i}}),$$
where $\kappa^{\theta_i,\theta_i}$ is the $i$th diagonal element of $K_n(\theta)^{-1}$ for
$i=1,2,3,4$ and $z_{\gamma/2}$ is the quantile $1-\gamma/2$ of the standard normal
distribution.

Further, we can compute the maximum values of the unrestricted and
restricted log-likelihoods to construct the LR statistics for
testing some sub-models of the BGE distribution. For example, we may
use the LR statistic to check if the fit using the BGE distribution
is statistically ``superior'' to a fit using the GE distribution for
a given data set. In any case, considering the partition
$\theta=(\theta_1^T,\theta_2^T)^T$, tests of hypotheses of the type
$H_0: \theta_1=\theta_1^{(0)}$ versus
$H_A:\theta_1\neq\theta_1^{(0)}$ can be performed by using any of
the above three asymptotically statistics. The LR statistic for
testing the null hypothesis $H_0$ is
$w=2\{\ell(\hat\theta)-\ell(\tilde\theta)\}$, where $\hat\theta$ and
$\tilde\theta$ are the MLEs of $\theta$ under $H_A$ and $H_0$,
respectively. Under the null hypothesis,
$w\stackrel{d}{\rightarrow}\chi_q^2$, where $q$ is the dimension of
the vector $\theta_1$ of interest. The LR test rejects $H_0$ if $w
>\xi_\gamma$, where $\xi_\gamma$ denotes the upper 100$\gamma$\% point of
the $\chi_q^2$ distribution.

\section{Application}

In this section we fit BGE model to one real data set. The data set is obtained
from Smith and Naylor \cite{Smith1987}. The data are the strengths of 1.5 cm glass fibres, measured at the National Physical Laboratory, England. Unfortunately, the units of measurement are not given in the paper. The data set is: 
0.55, 0.93, 1.25, 1.36, 1.49, 1.52, 1.58, 1.61, 1.64, 1.68, 1.73, 1.81, 2 ,0.74, 1.04, 1.27, 1.39, 1.49, 1.53, 1.59, 1.61, 1.66, 1.68, 1.76, 1.82, 2.01, 0.77, 1.11, 1.28, 1.42, 1.5, 1.54, 1.6, 1.62, 1.66, 1.69, 1.76, 1.84, 2.24, 0.81, 1.13, 1.29, 1.48, 1.5, 1.55, 1.61, 1.62, 1.66, 1.7, 1.77, 1.84, 0.84, 1.24, 1.3, 1.48, 1.51, 1.55, 1.61, 1.63, 1.67, 1.7, 1.78, 1.89.\\

\begin{figure}[h!]
\centering
\includegraphics[width=0.70\textwidth]{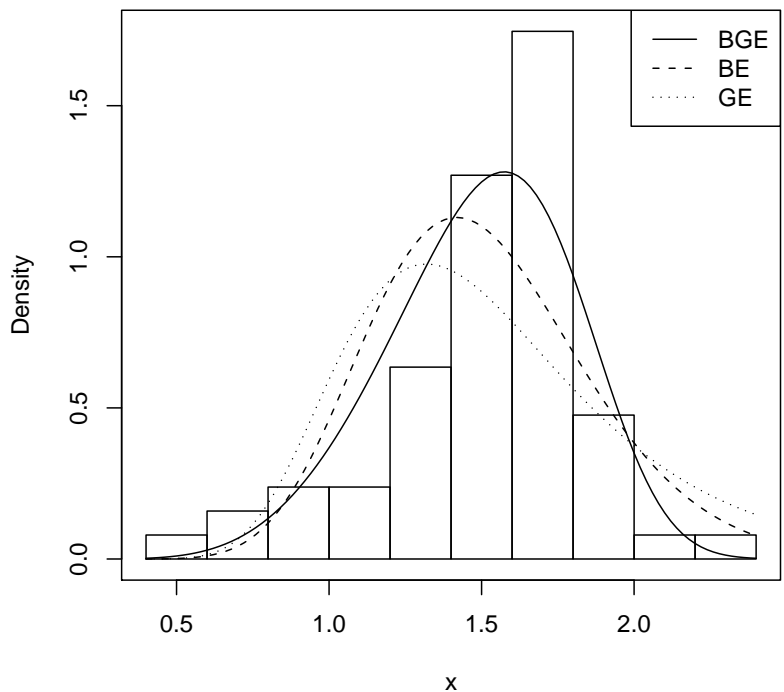}
\caption{Estimated densities of the BGE, BE and GE distributions for the data set from \cite{Smith1987}.}
\label{fig7}
\end{figure}

The MLEs of the parameters and the maximized log-likelihood for the BGE distribution are
\begin{eqnarray*}
\hat{a}=0.4125,\quad\hat{b}=93.4655,\quad\hat{\lambda}=0.92271,\quad\hat{\alpha}=22.6124,\quad
\hat{\ell}_{BGE}= -15.5995,
\end{eqnarray*}
whereas for the BE distribution
\begin{eqnarray*}
\hat{a}=17.7786,\quad\hat{b}=22.7222,\quad\hat{\lambda}=0.3898,
\quad\hat{\ell}_{BE}= -24.1270
\end{eqnarray*}
and for the GE distribution
\begin{eqnarray*}
\hat{\lambda}=2.6105, \quad\hat{\alpha}=31.3032,
\quad\hat{\ell}_{GE}=-31.3834.
\end{eqnarray*}
The LR statistics to test the hypotheses $H_0:BE\,\times\,H_A:BGE$
and $H_0:GE\,\times\,H_A:BGE$ are $17.0550$ 
(p-value=$3.63\times 10^{-5}$) and $31.5678$
(p-value=$1.39\times10^{-7}$), respectively. Therefore, we reject
the null hypothesis in both cases in favor of the BGE distribution
at the significance level of 5\%. The plots of the estimated
densities of the BGE, BE and GE distributions fitted to the data set
 given in Figure \ref{fig7} show that the BGE distribution gives a
better fit than the other two sub-models.

\section{Conclusions}
We proposed the beta generalized exponential (BGE) distribution
which ge\-ne\-ra\-li\-zes the beta exponential distribution discussed 
by Nadarajah and Kotz \cite{Nadarajah2005} and the generalized exponential 
(also named exponentiated exponential) distribution introduced by
Gupta and Kundu \cite{Gupta1999}. We provide a mathematical treatment of this
distribution including the densities of the order statistics. We derive
the moment ge\-ne\-ra\-ting function and provide infinite sums for the moments of the 
new distribution and of its order statistics. We discuss maximum likelihood 
estimation and obtain the information matrix. One application of the BGE distribution 
are given to show that this distribution could give better fit than other sub-models discussed
in the literature. We hope this generalization may attract wider
applications in reliability and biology.

\section*{Appendix}
We now give some details of (\ref{mgfbe}). It follows for $t<\lambda$ that\\
\begin{eqnarray*}
  B(b-t/\lambda,a)&=&B(a,b-t/\lambda)=\int_0^1\omega^{a-1}(1-\omega)^{b-t/\lambda}d\omega\\
 &=&\int_0^1\omega^{a-1}(1-\omega)^{1-t/\lambda}(1-\omega)^{b-1}d\omega.
\end{eqnarray*}
If $b>0$ is real non-integer, we obtain using (\ref{expansao}) in $(1-\omega)^{b-1}$
\begin{eqnarray*}
B(b-t/\lambda,a)&=&\sum_{j=0}^\infty\frac{(-1)^j\Gamma(b)}{\Gamma(b-j)j!}\int_0^1\omega^{a+j-1}(1-\omega)^{1-t/\lambda}d\omega\\
 &=&\sum_{j=0}^\infty\frac{(-1)^j\Gamma(b)}{\Gamma(b-j)j!}B(a+j,1-t/\lambda)\\
 &=&\sum_{j=0}^\infty\frac{(-1)^j\Gamma(b)}{\Gamma(b-j)j!}B(1-t/\lambda,a+j).
\end{eqnarray*}
Similarly for $b>0$ integer and using the binomial expansion in $(1-\omega)^{b-1}$, we have
\begin{eqnarray*}
B(b-t/\lambda,a)&=&\sum_{j=0}^{b-1}\binom{b-1}{j}B(1-t/\lambda,a+j).
\end{eqnarray*}
Hence, it is easy to see that (\ref{mgfbe}) holds.\\

Some useful quantities are defined by

\begin{eqnarray*}
   c_j&=&\psi(\alpha(a+j)+1)-\psi(1),\quad d_j=c_j^2+\psi'(1)-\psi'(\alpha(a+j)+1),\\
   e_j&=&-c_j[c_j^2+3\{\psi'(1)-\psi'(\alpha(a+j)+1)\}]+\psi''(1)-\psi''(\alpha(a+j)+1),\\
   f_j&=&\{c_j^2+\psi'(1)-\psi'(\alpha(a+j)+1)\}[c_j^2+3\{\psi'(1)-\psi(\alpha(a+j)+1)\}]+\\
   &&2c_j^2\{\psi'(1)-\psi'(\alpha(a+j)+1)\}-4c_j\{\psi''(1)-\psi''(\alpha(a+j)+1)\}.
\end{eqnarray*}

\end{document}